# SAT-Based Complete Don't-Care Computation for Network Optimization

**Alan Mishchenko and Robert K. Brayton**
Department of EECS
University of California, Berkeley
{alanmi, brayton}@eecs.berkeley.edu

## Abstract

*This paper describes an improved approach to Boolean network optimization using internal don't-cares. The improvements concern the type of don't-cares computed, their scope, and the computation method. Instead of the traditionally used compatible observability don't-cares (CODCs), we introduce and justify the use of complete don't-cares (CDC). To ensure the robustness of the don't-care computation for very large industrial networks, a optional windowing scheme is implemented that computes substantial subsets of the CDCs in reasonable time. Finally, we give a SAT-based don't-care computation algorithm that is more efficient than BDD-based algorithms. Experimental results confirm that these improvements work well in practice. Complete don't-cares allow for a reduction in the number of literals compared to the CODCs. Windowing guarantees robustness, even for very large benchmarks on which previous methods could not be applied. SAT reduces the runtime and enhances robustness, making don't-cares affordable for a variety of other Boolean methods applied to the network.*

## 1 Introduction

Optimization of Boolean networks using don't-cares plays an important role in technology-independent logic synthesis and incremental re-synthesis of mapped netlists. Traditionally, only satisfiability don't-cares (SDCs) and compatible observability don't-cares (CODCs) have been used [20]. The classical algorithm to compute CODCs [22] implemented in SIS [24] is used for many industrial tools. Later improvements dealt with (a) a more robust implementation [21], (b) independence from the local function representation [2], (c) generalization to multi-valued networks [7], and (d) approximations [23].

CODCs form a subset of the complete don't-cares (or complete flexibility) [12] projected onto a node by its context in the multi-level network. It was shown experimentally [11] that the computation of CDCs is comparable in runtime and memory requirements while, as expected, the amount of don't-cares computed is larger than for CODCs. The presentation in [11][12] considers the most general case of non-deterministic multi-valued networks, leaving open questions about efficiency when applied to binary deterministic networks.

The *first contribution of this paper* is in developing a specialized efficient version of the multi-valued don't-care computation algorithm [11][12], to work on binary networks, and in showing that, compared to CODCs, this algorithm leads to an increase in optimization quality, due to the additional freedom provided by the CDCs.

The *second contribution* concerns the computation of don't-cares in large industrial designs. The traditional don't-care optimization in SIS is performed using the whole network as the context for each node. This restricts the use of don't-cares to small or medium-sized networks. To apply the same method to larger networks, the network can be partitioned with the scope of computation limited to one partition at a time. Although not published, such methods are probably part of some industrial tools. However, we suspect that partitioning for don't care computation is difficult, ad hoc, and implementation dependent. We propose a non-partitioning scheme, called *windowing*, which efficiently trades quality for runtime in network optimization. Windowing captures the maximal flexibility within a context limited by a fixed number of logic levels surrounding the node in question. The reconvergent paths surrounding the node are included into the window and excluded as in [23]. Windowing is fast because construction of a window for a node involves a limited number of surrounding nodes visited without traversing the whole network. Windowing is not a partitioning scheme because each node has its own window that may overlap with windows computed for other nodes. Finally, windowing is dynamic and can be performed "on the fly", without duplicating or otherwise modifying the network or its parts. The latter quality makes windowing useful for applications that frequently update the network, e.g. decomposition-mapping [13].

The *third contribution* concerns the use of Boolean satisfiability [9][16] rather than BDDs or SOPs, for the computation of don't-cares. We show that SAT is responsible for a speed-up making CDCs easy to compute and affordable enough. As a result, many procedures that previously relied on algebraic methods can now be extended to use Boolean methods based on CDCs.

In combination, these contributions provide improved efficiency, quality, and ruggedness for technology-independent logic synthesis.

The paper is structured as follows: Section 2 establishes the background. Section 3 defines CDCs and compares them with CODCs. Section 4 presents windowing. Section 5 describes and compares BDD-based and SAT-based approaches to CDC computation. Section 6 gives experimental results, and Section 7 concludes the paper.

## 2 Background

**Definition**. A *completely specified Boolean function* (CSF) is a mapping from *n*-dimensional ($n \geq 0$) Boolean space into a single-dimensional one: $\{0,1\}^n \to \{0,1\}$.

A *don't-care* for a logic function allows it to have either 0 or 1 as a possible value. If for at least one input combination, the output of a function is a don't-care, this function is called an *incompletely specified Boolean function* (ISF).

An assignment of *n* Boolean variables to particular values is called a *minterm*. A CSF has *negative* (*positive*) *minterms* that correspond to the assignments, for which it takes values 0 (1). The positive and negative minterms are called the *care minterms*. An ISF has *don't-care minterms,* which correspond to the assignments where the function can be either 0 or 1.

A CSF is *compatible* with an ISF (*implements* the ISF), if the CSF can be derived from the ISF by assigning either 0 or 1 to each don't-care minterm. One ISF is said to be *larger* than another if it has more don't-care minterms.

**Definition**. A *Boolean network* is a directed acyclic graph (DAG) with nodes represented by Boolean functions. The sources of the graph are the *primary inputs* (PIs) of the network; the sinks are the *primary outputs* (POs).

The same name is used for a node and its output signal. The output of a node may be an input to other nodes called its *fanouts*. The inputs of a node are called its *fanins*. If there is a path from node *A* to *B*, then *A* is in the *transitive fanin* of *B* and *B* in the *transitive fanout* of *A*. The transitive fanin of *B*, *TFI*(*B*), includes node *B* and the nodes in its transitive fanin, including the PIs. The *transitive fanout* of *B*, *TFO*(*B*), includes node *B* and the nodes in its transitive fanout, including the POs.

The functionality of a node in terms of its immediate fanins is its *local function*. Its functionality in terms of the primary inputs of the network is its *global function*.

## 3 Complete don't-cares

Consider an individual node represented by a local CSF. It is not possible to change the node's function without changing the node's behavior. However, the situation is different when the node is considered in its context in the network. Then, often the node's function can be substantially modified without changing the behavior of the network. This is because other nodes prevent some combinations of inputs from reaching the node (non justification) as well as hiding the node's output from the POs under some conditions (non propagation).

The flexibility allowed in the implementation of a node can be represented as an ISF. A don't-care minterm of the ISF represents a combination of the node's input variables, for which the value of the node's output is not required for the POs of the network to produce the correct values.

**Definition**. The *complete don't-cares* (CDCs), or *complete flexibility* (CF), of a node in the binary network, is the largest ISF, whose don't-care minterms represent conditions when the output of the node does not influence the values produced by the POs of the network.

CDCs are important for network optimization because replacing a node's function by any CSF compatible with its CDC, does not change the network's output.

A key observation is that CDCs are *not* compatible, unlike CODCs [22]. That is, some POs of the network may produce incorrect values if CDCs are derived for several nodes and used independently. However, if a CDC is computed for a node and used immediately to optimize and replace that node before computing the CDC of another node, compatibility is not required. In this case, whenever a CDC is computed and used for a node, all prior changes to the other nodes are reflected in the computation. Heuristically, we found that visiting the nodes in topological order from the POs to the PIs gives the best literal reduction in a CDC-based optimization scheme.

CDCs have two major parts, the satisfiability don't-cares (SDCs) arising because some combinations are not produced as the inputs of the node, and the observability don't-cares (ODCs) arising because under some conditions the output value of the node does not matter. In Figure 1, node *F* has SDCs in the local space ($x = 0, y = 1$) due to limited controllability, while node *G* has ODCs in the global space ($a = 1, b = 1$) due to limited observability.

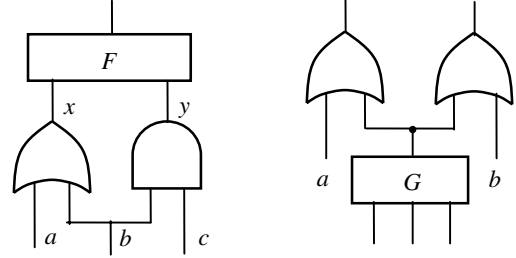

**Figure 1. Example of SDCs and ODCs.**

Don't-care computations are traditionally performed in the context of the entire Boolean network, as exemplified by SIS [24]. In the case of CDCs, this approach guarantees that the don't-cares computed are the largest set of don't-cares possible for a node. Since these computations can be expensive, we developed an optional *windowing* method that limits the scope of the don't-care computation to only a few logic levels on the fanin/fanout side of the node. A key observation is that re-convergence is a prime reason for don't cares. Therefore, along with the near TFI and TFO of a node, a window should contain all *re-convergent paths* that begin and terminate in these nodes. Previous approaches to windowing for don't-care computation [23] considers only the TFI and TFO of the node without considering re-convergence.

For the special case when the inputs to the window have disjoint supports in terms of the PIs and all outputs of the window are POs, the CDC computed for the window is equal to the CDC when the whole network is considered.

## 4 Windowing

This section contains a detailed discussion of the windowing algorithm introduced in [13].

**Definition**. Given a DAG and two non-overlapping subsets of nodes, one set is called the *leaf set* and the other the *root set*, if every path from the sources of the DAG to any node in the root set passes through some node in the leaf set.

**Definition**. Given two subsets in the leaf/root relationship, its *window* is the subset of nodes of the DAG that contains the roots plus all nodes on paths between the leaf set and the root set. The nodes in the leaf set are not included in the window.

**Definition.** A path between a pair of nodes is *distance-k* if it spans exactly *k* edges between the pair.

**Definition.** Two nodes are *distance-k* from each other if the shortest path between them is distance-*k*.

The pseudo-code in Figure 2 and the example in Figure 3 describe the flow of a window construction algorithm. Procedure *Window* takes a node and two integers that define the number of logic levels on the fanin/fanout sides of the node to be included in the window. It returns the leaf set and the root set of the window. With modifications, this procedure can compute a window for a *set* of nodes that, in general, need not be adjacent nor in the fanin/fanout relationship.

```
(nodeset, nodeset) Window( node N, int nFanins, int nFanouts )
{
    nodeset I₁ = CollectNodesTFI( {N}, nFanins );
    nodeset O₁ = CollectNodesTFO( {N}, nFanouts );
    nodeset I₂ = CollectNodesTFI( O₁, nFanins + nFanouts );
    nodeset O₂ = CollectNodesTFO( I₁, nFanins + nFanouts);
    nodeset S = I₂ ∩ O₂;
    nodeset L = CollectLeaves( S );
    nodeset R = CollectRoots( S );
    return (L, R);
}
```

**Figure 2. Computation of a window for a node.**

The procedure *CollectNodesTFI* takes a set $S$ of nodes and an integer $m \geq 0$, and returns a set of nodes on the fanin side that are distance-$m$ or less from the nodes in $S$. An efficient implementation of this procedure for small $m$ (for most applications, $m \leq 10$) iterates through the nodes that are distance-$k$ ($0 \leq k \leq m$) from the given set. The distance-$0$ nodes are the original nodes. The distance-$(k+1)$ nodes are found by collecting the fanins of the distance-$k$ nodes not visited before. The procedure *CollectNodesTFO* is similar.

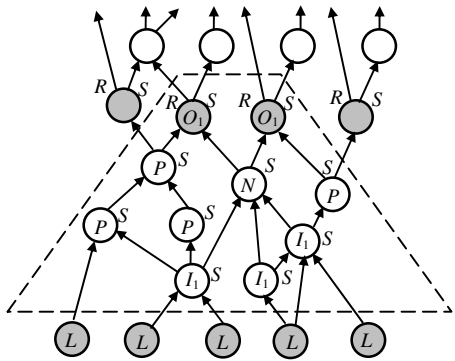

**Figure 3. Example of a 1 x 1 window.**

Procedures *CollectLeaves* and *CollectRoots* take the set of the window's internal nodes and determine the leaves and roots of this window. The leaves are the nodes that do not belong to the given set but are fanins of at least one of the nodes in the set. The roots are the nodes that belong to the given set and are also fanins of at least one node not in the set. Note that some of the roots thus computed are not in the TFO cone of the original node(s), for which the window is being computed, and therefore can be dropped without violating the definition of the window and undermining the usefulness of the window for the don't-care computation.

We typically refer to the window constructed for a node by including $n$ TFI logic levels and $m$ TFO logic levels as an $n \times m$ window.

**Example:** Figure 3 shows a $1 \times 1$ window for node $N$ in a DAG. The nodes labeled $I_1$, $O_1$, $S$, $L$, and $R$ are in correspondence with the pseudo-code in Figure 2. The window's roots (top) and leaves (bottom) are shaded. Note that the nodes labeled by $P$ do not belong to the TFI and TFO cones of node $N$, but represent the reconvergent paths in the vicinity of node $N$. The left-most root and right-most root are not in the TFO of $N$ and can be dropped, as explained above.

## 5 Don't-care computation

The network optimization discussed in this paper iterates through all the nodes of the network. For each node, a CDC is computed and used to simplify and replace the node before optimizing the next node. The computation of the CDC for a node can be performed in the context of the whole network, if the network is small; otherwise, a window is used. Without limiting the generality of the CDC computation, we discuss these methods as applied to a node in the whole network. If a window is used, the network is the sub-network defined by the window used for the node.

The general approach to computing the CDC of a node in a non-deterministic multi-valued network [11][12] relies on the use of an additional variable $z$ for the output of the node, and the computation of a Boolean relation in terms of $z$ and the PI and PO variables.

For a Boolean (binary deterministic) network, this approach can be simplified. The computation can be performed without using $z$ or Boolean relations. In both BDD-based and SAT-based implementations, we consider two instances of the same network that differ only in an inverter at the output of the given node in the second copy (Figure 4). This duplication is an imaginary construction, done for the sake of the presentation and not actually implemented in software.

The first network is the original one, while the second has an invertor inserted at the node's output. The functionalities of these networks are compared to detect when the change in the node's behavior influences the values at the POs. To this end, the two networks are transformed into a "miter" [1] derived by combining the pairs of PIs with the same names and feeding the pairs of POs with the same names into EXOR gates ORed to produce the only output of the miter (Figure 4).

### 5.1 Computation using BDDs

We use $x$ to represent the PIs of the network and $y$ to represent the immediate fanins of the node to be minimized. The BDD-based CDC computation begins by deriving the global functions of the POs of the two networks, $\{f_i(x)\}$ and $\{f_i'(x)\}$. Next, the output function of the miter, $C(x)$, is derived, which represents the care set in the global space:

$$C(x) = \sum_i [f_i(x) \oplus f_i'(x)].$$

The ODC of the node in the global space is the complement of the care set:

$$ODC(x) = \overline{C(x)} = \prod_i [f_i(x) \equiv f_i'(x)],$$

Next, the local CDC is computed by imaging the global ODC into the local space using the mapping $M(x,y)$ (inferred from the network) that relates the global and local spaces:

$$CDC(y) = \forall_x [M(x,y) \Rightarrow ODC(x)] = \forall_x [\overline{M(x,y)} + ODC(x)].$$

This computation adds the SDC, $\overline{M(x,y)}$, to the already computed ODC. Thus a don't-care minterm $y$ is, for all assignments of the PI variables $x$, either an SDC or an ODC. If external don't-cares are available, they are simply added to the ODC.

### 5.2 Computation using SAT

The use of SAT [9][16] in the CDC computation is similar to the use of SAT in combinational equivalence checking [4]. A solution of the SAT problem corresponding to the miter in Figure 4 gives a satisfying assignment for all network signals when a 1 is the output of the miter. The values of variables $y$ (the fanins of the node) in this solution form a *care set minterm* in the local space of the node. This is because, for them, we know there exist values of the PI variables *x*, such that at least one pair of POs produces different values.

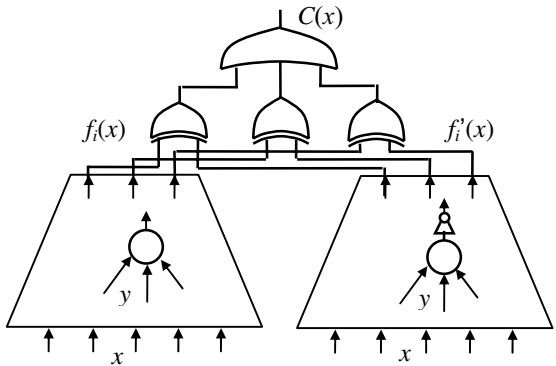

**Figure 4. Illustration of a miter used in CDC computation.**

All the care set minterms in terms of variables *y* are collected by enumerating through the satisfying assignments of the SAT problem and adding breaking clauses for each of them. A similar method of generating the satisfying assignments is described in [10], except that we do not undo the implication graph when a new satisfying assignment is found. We treat satisfying assignments similar to conflicts. In both cases, non-chronological backtracking is performed to the highest level determined using the new clause.

The SAT-based CDC computation is summarized in Figure 5. The top-level procedure *CompleteDC* takes node *N* and its context *S* given by the network, or by a window constructed for node *N*. Procedure *ConstructMiter* applies structural hashing [8] to the miter of the two copies of *S* shown in Figure 4. The resulting compact AND-INV graph *G* is constructed in one DFS traversal of the nodes in *S*, without actual duplication.

For efficiency, random simulation is used to derive part of the care set, $F_1$. The CNF *P* is the conjunction of clauses derived from *G* and the complement of $F_1$. The CNF of *G* is derived using a technique that adds three CNF clauses for each AND gates. For example, the clauses added for the gate $ab = c$ are: $\overline{c} + a$, $\overline{c} + b$, $\overline{a} + \overline{b} + c$. The only other clause added to the CNF is the clause asserting that the PO of the miter is equal to 1.

The SAT solver enumerates through the satisfying solutions, $F_2$, of the resulting problem representing the remaining part of the care set. In practice, it often happens that the SAT problem has no solutions ($F_2 = 0$). In such cases, SAT is only useful to prove the completeness of the care set derived by random simulation.

```
function CompleteDC( node N , context S )
{
    aig G = ConstructMiter( S, N );
    function F₁ = RandomSimulation( G );
    cnf P = CircuitToCNF( G ) ∧ FunctionToCNF( F̄₁ );
    function F₂ = SatSolutions( P );
    return F₁ + F₂ ;
}
```
**Figure 5. Pseudo-code of SAT-based CDC computation.**

This approach solves the SAT problem by enumerating through the satisfying assignments that represent local *minterms* of the care set of the given node. Therefore, it should be limited to nodes with roughly 10 inputs or less, which is typically the case for most Boolean networks. It could also be enforced by decomposing large nodes first. To make the approach appropriate for networks nodes with a larger number of inputs, the implementation of the SAT solver should be further modified to return incomplete satisfying assignments corresponding to *cubes* rather than minterms of the local care set.

## 6 Experimental results

The methods for computing CDCs of a node in the context of both a window and the whole network have been implemented in the MVSIS environment [19].

The SAT-based part was implemented using MiniSat [3], an "extensible SAT solver". Despite its small size (600 lines of C++ code written without STL), MiniSat is very efficient. In our experiments, it outperformed several popular SAT solvers. Moreover, the implementation of MiniSat is easy to understand and modify, which was the original intention of its developers.

Two experiments were performed. In both cases, the measurements were done on a Windows XP computer with a 1.6GHz CPU and 1Gb RAM, although less than 256Mb of RAM are needed for the largest benchmarks in Table 2.

The resulting networks were verified using a SAT-based verifier in MVSIS designed along the lines of [4][6].

### 6.1 Experiment 1: Comparing CODCs vs. CDCs

We compared the optimization potential of CODCs and CDCs. The BDD-based don't-care computation flow was used in both cases. We considered the largest MCNC benchmarks [25], for which BDDs could be constructed. Table 1 compares the runtime and number of literals produced by the CODC-based command *full_simplify* of SIS, and the new CDC-based command *mfs* implemented in MVSIS and later ported to SIS. The SIS version was used in this experiment. Both *full_simplify* and *mfs* perform Boolean resubstitution followed by SOP minimization as part of a don't-care-based optimization. Network *sweep* in SIS is performed before and after both commands.

The first column in Table 1 lists the benchmark names, followed by five columns containing the number of literals: (1) after initial sweeping only ("sweep") (which is the starting point of the other columns), (2) after *full_simplify* ("fs"), (3) after *mfs* without the "advanced features" ("mfs"), (4) after *mfs* with 2 x 2 windowing without the "advanced features" ("mfsw"), and (5) after *mfs* with the advanced features enabled ("MFS"). The advanced features include on-the-fly merging of nodes with functionality equivalence up to complementation and phase-assignment, performed as part of optimization. In columns (2) and (3) these features are disabled to have a fair comparison with *full_simplify*. Some benchmarks could not be processed by *full_simplify* because of the large BDD sizes (indicated by the dash in the table).

The last three columns give the runtimes in seconds. The bottom line shows the average of the ratios of the improvements in the number of literals, achieved by each command, compared to the number of literals in the original (swept) benchmarks. The asterisk in Table 1 indicates that, to compare against *fs*, the averages of the ratios are taken only over the 11 examples where *fs* could complete.

## Table 1. Comparing CODCs vs. CDCs.

| Name | Literals in factored forms | | | | | Runtime, sec | | |
|---|---|---|---|---|---|---|---|---|
| | sweep | fs | mfs | mfsw | MFS | fs | mfs | mfsw |
| dalu | 2976 | 2140 | 1741 | 2250 | 1747 | 64.8 | 2.1 | 0.8 |
| des | 6101 | 5677 | 5616 | 5920 | 5334 | 8.1 | 3.7 | 3.7 |
| frg2 | 2010 | 1454 | 1440 | 1477 | 1409 | 5.1 | 0.6 | 0.5 |
| i10 | 4355 | - | 3809 | 3853 | 3694 | - | 82.2 | 1.2 |
| k2 | 2928 | 2889 | 2663 | 2878 | 2641 | 6.2 | 3.9 | 3.3 |
| pair | 2420 | 2179 | 2143 | 2151 | 2139 | 3.5 | 2.9 | 0.4 |
| c1355 | 992 | 984 | 992 | 992 | 992 | 22.8 | 86.7 | 0.2 |
| c1908 | 1058 | 869 | 870 | 869 | 754 | 12.4 | 10.9 | 0.3 |
| c2670 | 1570 | 1189 | 1215 | 1411 | 1195 | 4.9 | 2.8 | 0.3 |
| c432 | 335 | 298 | 288 | 299 | 288 | 2.2 | 0.9 | 0.3 |
| c499 | 576 | 568 | 576 | 576 | 576 | 1.0 | 13.0 | 0.1 |
| c5315 | 3531 | 3184 | 3168 | 3176 | 2951 | 31.5 | 7.3 | 0.9 |
| c7552 | 4750 | - | 4057 | 4079 | 3594 | - | 50.0 | 1.4 |
| c880 | 648 | 625 | 624 | 625 | 624 | 1.2 | 7.2 | 0.1 |
| **Ave** | **1.00** | **0.88*** | **0.86** | **0.90** | **0.83** | **1.00** | **0.87** | **0.07** |

Although *full_simplify* was expected to be faster than *mfs*, this was not the case possibly because of the differences in the implementation and use of different BDD variable ordering heuristics in SIS and MVSIS. Comparing literals, Table 1 shows that the CDCs outperform CODCs in the context of the whole network (columns "fs" vs. "mfs"). In those rare cases when CODCs give better literal counts, the improvement is attributed to finding a better ordering of nodes. Our experiments have shown that, on average over all considered benchmarks, CDCs typically contain 20% more don't-care minterms in the local spaces of the nodes, compared to the CODCs.

For CDCs with windowing (column "mfsw"), Table 1 shows that the literal count is close to that of CODCs in the context of the whole network (column "fs"), but the runtime is only 7% of that of "fs". Thus, the improvement due to the CDCs is only marginally outweighed by the degradation due to using a window instead of the whole network in the case of CODCs. Additionally, window-based optimization (*mfsw*) is applicable to very large circuits, well beyond the scope of *full_simplify* in SIS or *mfs* <u>without windowing</u>.

### 6.2 Experiment 2: Cumulative effect of improvements

Table 2 shows the results of network optimization using the SAT-based flow for ITC'99 [5], ISCAS, and PicoJava benchmarks [17]. These benchmarks are relatively large. As a result, BDD-based methods, *full_simplify* in SIS and *mfs* in MVSIS without windowing, cannot be applied.

The first column of Table 2 lists the benchmark names. The second column shows the number of inputs, outputs, and latches. The next three columns contain the number of literals in the factored forms in (1) the original benchmark after sweeping ("sweep"), (2) after applying *mfs* with 2x2 windowing ("mfsw"), and (3) as part of a script ("script"). The last two columns show the runtime in seconds for *mfsw* and *script*. The script used in this experiment was *mvsis.rugged* which is similar to *script.rugged* of SIS, except that *mvsis.rugged* is implemented in MVSIS and the CODC-based SIS command *full_simplify* is replaced by the CDC-based MVSIS command *mfs*, using 2x2 windows (*mfs –w* 22) and SAT instead of BDDs.

## Table 2. Network optimization using CDCs, windowing and SAT.

| Name | In/Out/Latch | Literals in factored forms | | | Runtime, s | |
|---|---|---|---|---|---|---|
| | | sweep | mfsw | script | mfsw | script |
| b14 | 32 / 54 / 245 | 17388 | 10664 | 7911 | 3.9 | 18.0 |
| b15 | 36 / 70 / 449 | 16244 | 15056 | 10948 | 6.1 | 22.9 |
| b17 | 37 / 97 / 1415 | 57311 | 49067 | 37877 | 35.7 | 104.8 |
| b20 | 32 / 22 / 490 | 35149 | 21826 | 16813 | 7.6 | 55.0 |
| b21 | 32 / 22 / 490 | 35908 | 22312 | 16932 | 9.3 | 51.1 |
| b22 | 32 / 22 / 735 | 52276 | 33017 | 25174 | 13.5 | 59.8 |
| s15850 | 14 / 87 / 597 | 7303 | 6350 | 4033 | 1.2 | 4.0 |
| s35932 | 35 / 320 | 24408 | 20248 | 10986 | 4.2 | 16.7 |
| s38417 | 28 / 106 | 18699 | 17327 | 13640 | 4.5 | 15.5 |
| pj1 | 1769 / 1063/0 | 34828 | 30547 | 18076 | 9.5 | 37.0 |
| pj2 | 690 / 429/0 | 7422 | 6464 | 3457 | 1.1 | 4.0 |
| **Ave** | | **1.00** | **0.79** | **0.54** | **1.00** | **4.36** |

Table 2 shows that the proposed don't-care-based optimization flow can be applied to quite large circuits. This is because the don't-care computation is performed in a window, and therefore is local and does not depend on the circuit size. The overall runtimes scale well with the problem size and are predictable; a rule of the thumb is for *mfs –w* 22, the computation takes about 1 second per 3000 literals.

The reader can refer to a workshop version of the paper [14] for the detailed comparison of the performance of SAT and BDDs with windows of different sizes. In summary, the SAT-based computations are faster and scale better than the BDD-based ones. Thus, for 1×1 windows, SAT is on average 20% faster; for 2×2 windows, it is over two times faster while for 4×4 windows, it is over 7 times faster. This ratio increases further with the window size.

## Acknowledgements

The authors gratefully acknowledge the support of the California MICRO program and our industrial sponsors, Intel, Fujitsu, Magma, and Synplicity.

## 7 Conclusions

This paper contributes several improvements to the optimization of logic networks using don't-cares:
- Complete don't-cares are used instead of compatible don't-cares. Abandoning compatibility does not lead to any problems in runtime but does increase the amount of don't-cares computed.
- To ensure robust computation of don't-cares, windowing is used. This technique noticeably reduces the runtime while computing a substantial subset of the complete don't-cares for each node.
- A new implementation of the don't-care computation using Boolean satisfiability is used, taking advantage of the recent improvements in the performance of SAT solvers [16]. The same set of don't-cares is computed as in the corresponding BDD-based algorithm, but several times faster.

The experiments described in the paper show that the proposed improvements enhance the optimization quality reduce the runtime and provide robustness. The overall effect is that the computation of internal don't-cares becomes very affordable, even for very large industrial networks.

Since such ideas make CDC computations quite affordable and robust, we believe that they can be applied to other Boolean logic optimization methods, reducing computational cost and improving optimality. As a result, these Boolean methods become more affordable and should eventually replace some sub-optimal algebraic methods for a variety of tasks in logic synthesis.